\documentclass[graybox]{svmult}

\usepackage{mathptmx}       
\usepackage{helvet}         
\usepackage{courier}        
\usepackage{type1cm}        
\usepackage{makeidx}         
\usepackage{graphicx}        
\usepackage{multicol}        
\usepackage[bottom]{footmisc}

\makeindex             

\begin{document}

\title*{Chandra observations of Cygnus~OB2}
\author{Nicholas J. Wright, Jeremy J. Drake, Janet E. Drew, Jorick S. Vink}
\authorrunning{Nicholas J. Wright}
\institute{Nicholas J. Wright \at Harvard-Smithsonian Center for Astrophysics, 60 Garden Street, Cambridge, MA 02138, USA. \email{nwright@cfa.harvard.edu}
\and Jeremy J. Drake \at Harvard-Smithsonian Center for Astrophysics, 60 Garden Street, Cambridge, MA 02138, USA.
\and Janet E. Drew \at Centre for Astronomy Research, University of Hertfordshire, Hatfield AL10 9AB, UK.
\and Jorick S. Vink \at Armagh Observatory, College Hill, Armagh BT61 9DG, UK.}

\maketitle

\abstract{Cygnus~OB2 is the nearest example of a massive star forming region, containing over 50 O-type stars and hundreds of B-type stars. We have analyzed two {\it Chandra} pointings in Cyg~OB2, detecting $\sim$1700 X-ray sources, of which $\sim$1450 are thought to be members of the association. Optical and near-IR photometry has been obtained for $\sim$90\% of these sources from recent deep Galactic plane surveys. We have performed isochrone fits to the near-IR color-magnitude diagram, deriving ages of $3.5^{+0.75}_{-1.0}$ and $5.25^{+1.5}_{-1.0}$~Myrs for sources in the two fields, both with considerable spreads around the pre-MS isochrones. The presence of a second population in the region, somewhat older than the present-day O-type stars, has been suggested by other authors and fits with the ages derived here. The fraction of sources with inner circumstellar disks (as traced by the $K$-band excess) is found to be very low, but appropriate for a population of age $\sim$5~Myrs. We measure the stellar mass functions and find a power-law slope of $\Gamma = -1.09 \pm 0.13$, in good agreement with the global mean value estimated by Kroupa. A steepening of the mass function at high masses is observed and we suggest this is due to the presence of the previous generation of stars that have lost their most massive members. Finally, combining our mass function and an estimate of the radial density profile of the association suggests a total mass of Cyg~OB2 of $\sim 3 \times 10^4$~M$_{\odot}$, similar to that of many of our Galaxy's most massive star forming regions.}

\section{Introduction}

Star formation occurs on scales ranging from small star forming regions (SFRs) to the massive superstar clusters seen in merging galaxies, yet our current understanding of star formation is mainly derived from observations of nearby small star forming regions, yet the conditions in these low-mass regions are very different to those found in the massive SFRs that contain hundreds to thousands of OB stars and millions of low-mass stars. High stellar densities, strong stellar winds from OB stars, and a large UV flux are likely to influence the products of the star formation process such as the initial mass function (IMF), the binary fraction, and the properties of proto-planetary disks. Unfortunately the majority of massive SFRs are found at great distances making them difficult to study. The exception to this is Cygnus~OB2, which at a distance of 1.45~kpc [1] is the closest known massive SFR, containing $\sim$65 O-type stars and thousands of OB stars [1,2,3,4]. However, the complications of high extinction, a large spatial area, and the inevitable foreground contamination caused by observing in the Galactic Plane have hindered previous attempts to study the low-mass population in the region. To overcome this, we have used {\it Chandra} observations to separate the X-ray luminous pre-MS population from the contaminating galactic foreground. Combining these observations with optical and near-IR photometry from recent deep photometric surveys of the Galactic plane we have been able to isolate the pre-MS population in the region.

\section{Cygnus~OB2 membership and foreground contamination}

\begin{figure}[t]
\sidecaption
\includegraphics[width=112.5pt, angle=270]{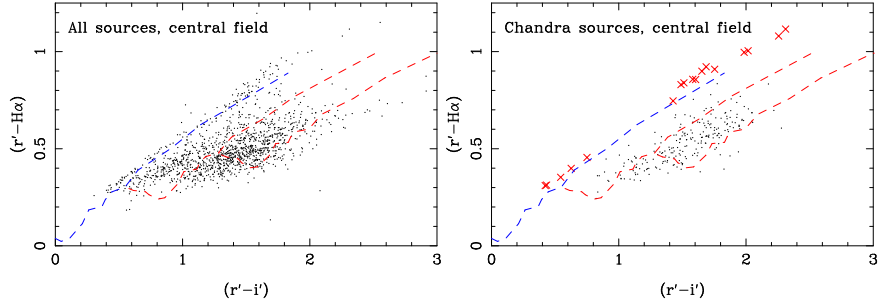}
\caption{IPHAS ($r' - $H$\alpha$, $r' - i'$) CCDs for all sources in the {\it Chandra} observational footprints (left) and for sources with {\it Chandra} associations (right). Blue lines show unreddened main-sequences [6], while red lines are the main-sequence tracks for $A_V = 4.0$ and 7.0. Sources removed from the catalogue as foreground sources are shown as red crosses.}
\label{foreground_iphas_ccd}
\end{figure}

Wright \& Drake [5] presented a catalogue of 1696 X-ray sources extracted from {\it Chandra} observations of two fields in the center of Cyg~OB2. The catalogue also includes optical and near-IR photometry from the IPHAS [6], 2MASS [7] and UKIDSS [8] surveys. X-ray selection is highly effective for selecting pre-MS stars, but to remove any remaining foreground contaminants in the catalog we employed the IPHAS ($r' - $H$\alpha$, $r' - i'$) color-color diagram (CCD, Figure~\ref{foreground_iphas_ccd}) where the main sequence does not redden onto itself but sweeps out an area in the color-color plane (e.g. [6,9]), allowing differently reddened stellar populations to be separated. Employing this method we removed 46 sources ($\sim$3\%), leaving a catalog of 1455 high-confidence members of Cyg~OB2 that Wright et al. [10] showed was complete to $\sim$1~M$_{\odot}$ (excluding late B and A-type stars that do not all emit X-rays).

\section{Near-IR stellar properties}

\begin{figure}[t]
\sidecaption
\includegraphics[width=190pt, angle=270]{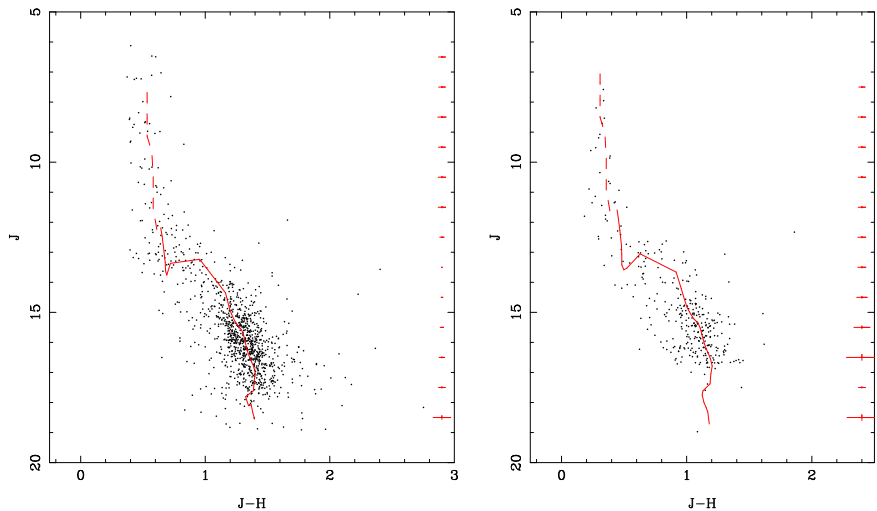}
\caption{Near-IR CMDs for Cygnus~OB2 sources in the central (left) and north-western (right) fields with pre-MS isochrones (red lines, [11]) for ages and extinctions of 3.5 and 5.25~Myrs, and $A_V = 7.5$ and $A_V = 5.5$, respectively. Mean photometric errors for each magnitude interval are shown as red crossed on the right.}
\label{nearir_cmd}
\end{figure}

The integrated properties of the region were studied in the near-IR due to the availability of deep near-IR photometry for the majority of sources. Figure~\ref{nearir_cmd} shows ($J$, $J-H$) color-magnitude diagrams (CMDs) for sources in the two fields. Pre-MS isochrones [11] were fit to this data to determine the age and extinction using a simple goodness-of-fit test and weighting each source using its color and magnitude. A bootstrapping technique was used to determine 95\% confidence intervals for all fitted quantities. We find different ages for the two fields of $3.5^{+0.75}_{-1.0}$ and $5.25^{+1.5}_{-1.0}$~Myrs for the central and north-western fields, respectively. However, the difference between these two ages is within the uncertainties of the two fits.

\begin{figure}[t]
\sidecaption
\includegraphics[width=210pt, angle=270]{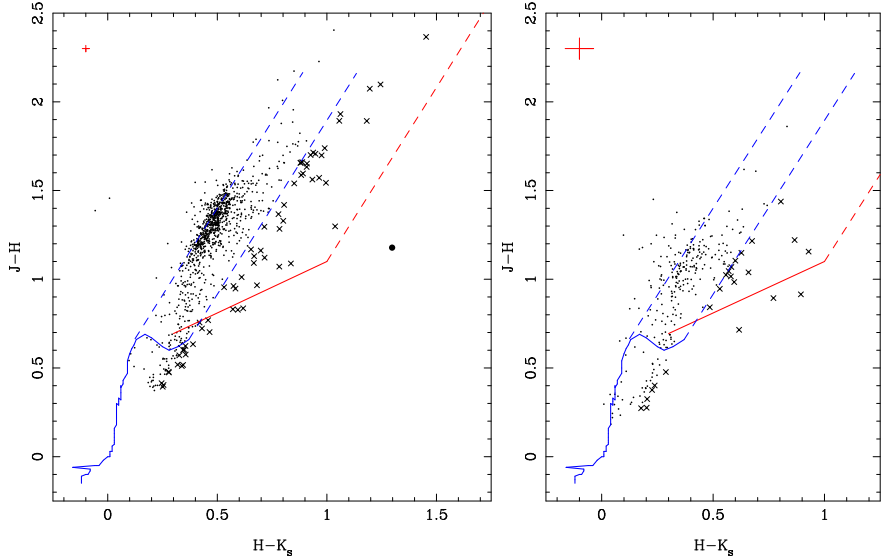}
\caption{Near-IR CCDs for Cygnus~OB2 sources in the central (left) and north-western (right) fields with a main sequence (blue lines, [15]), the classical T-Tauri star locus (red line, [16]), and reddening vectors (dashed lines) shown. Sources falling within the T-Tauri region are marked as crosses and the most highly reddened object in the field is shown as a large dot. The mean photometric errors on the two colors is shown in the top left corner of each figure in red.}
\label{nearir_ccd}
\end{figure}

Figure~\ref{nearir_ccd} shows the ($J-H$, $H-K_s$) CCD for the two fields, which was used to identify stars with circumstellar disks based on their near-IR excesses (in the absence of longer-wavelength data). We find 63 and 23 sources in this region of the color-color plane, 5.9\% and 7.9\% of our catalogue in the central and north-western fields respectively. These fractions are significantly lower than that found from near-IR studies of other similar age star forming regions [12], which has previously been suggested as evidence for the impact of environment, particularly the strong UV flux, on the evolution of protoplanetary disks [13]. While our results do support such a theory, with the fraction of circumstellar disks lower in the center of the star forming region where the UV radiation field from OB stars is stronger, the difference is both statistically small and over-simplifies the dynamical and 3-dimensional structure of the region. An alternative explanation is that we are sampling an older generation of star formation as well as the 2~Myr population. A disk fraction of 5~--~10\% is perfectly reasonable for a population with an age of $\sim$5~Myr [12].

\section{The stellar mass function in Cygnus~OB2}

\begin{figure}[t]
\sidecaption
\includegraphics[width=130pt, angle=270]{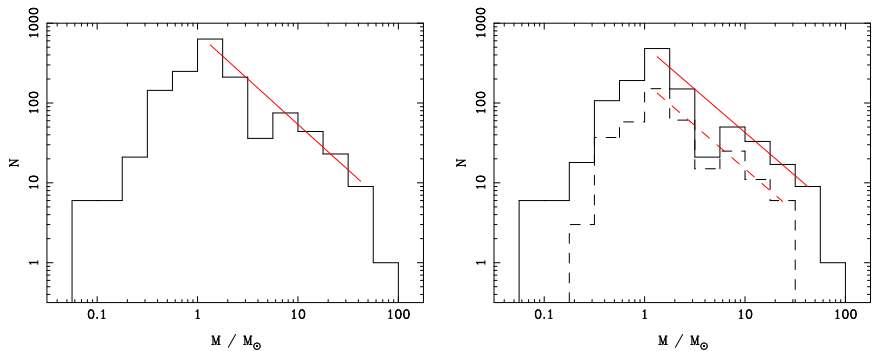}
\caption{Mass functions for all sources in Cygnus~OB2 with stellar masses derived from near-IR photometry and existing spectroscopic observations. {\it Left:} MF for both fields combined; {\it right:} MF for the central (full line) and north-western (dashed line) fields. $\chi^2$ fits to the MFs (within the completeness limits) are shown as red lines.}
\label{imf}
\end{figure}

Figure~\ref{imf} shows mass functions (MF) for sources in Cyg~OB2, exhibiting a clear power law slope (excluding A and late B-type stars) down to our completeness limit of $\sim$1~M$_{\odot}$, but steepening at higher masses. We fit a slope of $\Gamma = -1.09 \pm 0.13$ in the combined fields, in agreement with the apparently universal value of $\Gamma = -1.3$ [14]. When the two fields are considered separately we find negligibly different slopes of $\Gamma = -1.08 \pm 0.15$ in the central field and $\Gamma = -1.09 \pm 0.10$ in the north-western field. Previous measurements of the MF in Cyg~OB2 have mostly been derived from spectroscopy of the massive stars, the most recent measurement by Kiminki et al. [4] resulting in a slope of $\Gamma = -2.2 \pm 0.1$. We find here that the MF is not as steep as this, but confining our sample only to the O-type stars we find a slope of $\Gamma = -2.72 \pm 0.52$, in agreement with this work. We suggest that the steep high-mass slope of the MF is not intrinsic but is a product of the older stellar populations in the region having lost their most massive members, thereby steepening the MF.

\section{The age, star formation history, and total mass of Cygnus~OB2}

The main constraints on the age of Cyg~OB2 have previously been derived from the massive stars in the region. Hanson [1] estimated an age of $2 \pm 1$~Myrs but noted that as the search for OB stars in the region was extended to larger radii the fraction of evolved massive stars increased, contaminating the sample of Cygnus~OB2 with non-members. Many of these stars were found to the south of the original association [3] in a region where Drew et al. [9] have since uncovered a population of A-type dwarf stars with an age of 5--7~Myr. We have derived ages of $3.5^{+0.75}_{-1.0}$ and $5.25^{+1.5}_{-1.0}$~Myrs for the central and north-western fields, which could suggest a significant age spread across the Cygnus~OB2 association. Furthermore, our findings of an age-steepened mass function and a considerably reduced circumstellar disk fraction lend weight to the argument of a previous generation of stars in the region. In such a picture our central field includes the young OB-star dominated association, while the north-western field also includes older stars from a previous generation of star formation. We cannot yet know to what extent there are two distinct generations of stars present, or whether there is a real age spread.

While it is likely that Cygnus~OB2 is far from spherically symmetric, we can use a simple King profile to estimate the total size and mass of the association. Fitting a radial profile to our observations and combining it with our mass function (extrapolated down to 0.01~M$_{\odot}$ using a multi-stage power-law IMF [14]) we estimate a total stellar mass of $(3 \pm 1) \times 10^4$~M$_{\odot}$ for the entire association. Such a mass implies that there were originally $\sim$75 O-type stars, slightly larger than the currently known number of 65, but supporting our argument that some of the most massive stars have expired. This size is comparable to that of many of the most massive known star forming regions in our Galaxy, confirming the role of Cygnus~OB2 as the prime location to study the influence of massive stars on the low-mass stellar population.

\section{Conclusions}

We present an analysis of $\sim$1450 stellar X-ray sources detected in the Cygnus~OB2 association using complementary optical and near-IR photometry. We derive integrated ages, near-IR disk fractions and stellar masses, from which me measure the mass function. Evidence for an age spread, a low disk fraction, and an steep, high-mass slope of the mass function all argue for the presence of an older population of stars in the region. Finally we combine our results to estimate the total mass of the association as $3 \times 10^4$~M$_{\odot}$, comparable to many of our galaxy's most massive star forming regions. The recently completed {\it Chandra} Legacy Survey of Cygnus~OB2 will take this work wider and deeper and will help probe the star formation process in our nearest massive star forming region.

\end{document}